\documentclass[onecolumn, draftcls, 12pt]{IEEEtran}
\usepackage{amsfonts}
\usepackage{mathrsfs}
\usepackage{amssymb,amsmath}
\usepackage[noblocks]{authblk}
\usepackage[compress]{cite}
\usepackage{bm}
\usepackage{algorithm}
\usepackage{algorithmic}
\usepackage{amsmath,amssymb,epsfig,graphics,subfigure}
\usepackage{theorem}
\usepackage{array,color}
\usepackage[compress]{cite}
\usepackage{algorithm}
\usepackage{algorithmic}
\usepackage{color}
\def\DDcom#1{{\color{black} #1}}

\theoremheaderfont{\normalfont\bfseries}
\begin{document}

\title{Adaptive Multi-objective Optimization for Energy Efficient Interference Coordination in Multi-Cell Networks}

\author{Zesong Fei$^*$, Chengwen Xing, Na Li, and Jingming Kuang\\
 \thanks{ The authors are with School of Information and Electronics,
Beijing Institute of Technology, Beijing 100081, China.\\ E-mail:
feizesong@bit.edu.cn, chengwenxing@ieee.org, tsingnana@gmail.com,
jmkuang@bit.edu.cn.} \thanks{$^*$The corresponding author is Zesong
Fei.} }

\maketitle

\begin{abstract}
In this paper, we investigate the distributed power allocation for
multi-cell OFDMA networks taking both energy efficiency and
inter-cell interference (ICI) mitigation into account. A performance
metric termed as throughput contribution is exploited to measure how
ICI is effectively coordinated. To achieve a distributed power
allocation scheme for each base station (BS), the throughput
contribution of each BS to the network is first given based on a
pricing mechanism. Different from existing works, a bi-objective
problem is formulated based on multi-objective optimization theory,
which aims at maximizing the throughput contribution of the BS to
the network and minimizing its total power consumption at the same
time. Using the method of Pascoletti and Serafini scalarization, the
relationship between the varying parameters and minimal solutions is
revealed. Furthermore, to exploit the relationship an algorithm is
proposed based on which all the solutions on the boundary of the
efficient set can be achieved by adaptively adjusting the involved
parameters. With the obtained solution set, the decision maker has
more choices on power allocation schemes in terms of both energy
consumption and throughput. Finally, the performance of the
algorithm is assessed by the simulation results.
\end{abstract}

\newpage

\section{Introduction}

Ever-increasing demand on high data rates inspires and promotes the
development of wireless technologies. In order to achieve the
desired high information rates, many innovative ideas are introduced
into wireless system designs. Referring to cellular networks,
interference becomes the main limitation of performance improvement
and multi-point coordination is a promising and powerful technology
to improve the efficiency and reliability of wireless
communications. In the scenario of multi-cell orthogonal frequency
division multiple access (OFDMA) networks, a large number of users
try to share the same spectrum to carry out wide-band multimedia
communications and thus the performance of wireless networks is
heavily limited by mutual interference. This fact motivates
researchers to design various power control optimization algorithms
to effectively coordinate interference.

In a multi-cell network, generally speaking the design of
interference coordination is very challenging due to many practical
limitations \cite{Boudreau,Fodor,FeiGao}. One of the difficulties
for interference mitigation stems from the competition of utility
benefits among different base stations (BSs). Specifically, the
interactions between different BSs will greatly affect the whole
network performance. Hence, to efficiently schedule the inter-cell
interference the interactions among different BSs and the
characteristics of the BSs' behaviors should be carefully exploited.
In literature a successful model for the problem of interference
coordination is game theory which can effectively analyze the
behaviors of wireless nodes \cite{Lasaulce,Sung,Menon,Liang,Huang}.

In general, the game models exploited for wireless designs can be
classified into two categories: cooperative and non-cooperative
games. For non-cooperative game, it is convenient to devise totally
distributed algorithms, however it may suffer from a significant
performance loss compared with the optimal centralized solution due
to the fact that there is no cooperation among cells. On the other
hand, cooperative game usually suffers a cost of high overhead and
complexity though it has a better performance.

To overcome the inefficiency of the non-cooperative game and high
overhead of the cooperative game, recently pricing mechanism has
been proposed \cite{Schmidt}, which is employed as an effective
means to stimulate cooperation among players. Specifically, pricing
schemes can guide the players' behaviors toward efficient Nash
equilibrium, by introducing a certain degree of coordination in a
non-cooperative game. This approach has been introduced by C. Shi in
\cite{Non_Separable}. An algorithm for allocating power among
multiple interfering transmitters in a wireless network using OFDM
was presented. The algorithm attempts to maximize the sum over user
utilities, where each user's utility is a function of his total
transmission rate. Users exchange interference prices reflecting the
marginal cost of interference on each sub-channel, and then update
their power allocations given the interference prices and their own
channel conditions. Similar works have also been done in
\cite{OFDMA,MIMO,EW,beamforming,Beamformer} with local interference
pricing, and the corresponding algorithms adjusting beamforming
vectors or power allocation schemes to maximize the sum transmission
rate, respectively. It is interesting that though only limited
information is exchanged in these existing algorithms, the
performance of the distributed algorithms proved to be close to the
centralized optimization under a pricing mechanism
\cite{unified_approach}.

Game theory has achieved a great success in modeling the resource
allocation in multi-cell networks. In addition, it has also
introduced a series of successful mechanisms such as pricing
mechanism. Also inspired by the pricing mechanism, we take a further
step to tackle the resource allocation from the viewpoint of
multi-objective optimization theory. In our work, the performance
metrics of both throughput contribution and power consumption are
taken into account. In an interference-dominated network, at high
signal-to-noise ratio (SNR) regime, it is well-known that increasing
transmit power will be useless to improve system performance. On the
other hand, power consumption is also a very important performance
metric which should be carefully addressed in wireless network
designs, as green communications are of great importance in
practical applications. These two design issues are closely related
with each other, while for most of the existing works, they are
investigated separately. To the best of the authors' knowledge,
there are few works jointly considering energy efficiency and
mitigating the inter-cell interference at meantime
\cite{Fapojuwo,Choi}.

In this paper, based on a multi-objective optimization framework an
energy-efficient power optimization for multi-cell networks is
investigated. Using the pricing mechanism, a bi-objective
optimization problem is formulated for each BS, which aims at both
maximizing the throughput contribution of the BS to the network and
minimizing its total power consumption at the same time. The
throughput contribution of the BS is a function of the transmission
rate of the BS and his interference cost. To find out the tradeoff
between the throughput contribution of the BS and its power
consumption, the Pascoletti and Serafini scalarization method
\cite{mopsp} instead of the widely used weighted sum method is first
applied to transform the bi-objective problem into an equivalent
scalar objective problem considering that Pascoletti and Serafini
scalarization method has strengths in overcoming the deficiencies of
the weighted sum method. Then a proposed adaptive parameter control
algorithm is used to adaptively changing some parameters such that
the complete minimal solution set for the bi-objective optimization
problem can be derived. The minimal solutions are those on the
boundary of the set of feasible solutions. These points have a
common distinguishing characteristic. For a given constant power
consumption, the throughput contribution of the BS is the maximum
one and otherwise for a fixed throughput contribution, the consumed
power is the minimum. Finally, the performance advantage of the
proposed algorithm is demonstrated by the simulation results and it
shows that the proposed algorithm can effectively achieve different
tradeoffs among the involved objective functions. In addition
several famous solutions can be covered by our solution, such as the
solution to capacity maximization problem, the solution to
pricing-based utility optimization problem, and the solution to
equal power allocation problem.

This paper is organized as follows. In Section 2, the system model
is introduced and the corresponding multi-objective optimization
problem is formulated. The proposed adaptive parameter control
algorithm is detailed discussed in Section 3, in which we study the
connection between parameters and minimal solutions. Later the
performance of the algorithm is shown by simulations in Section 4.
Finally, conclusions are drawn in Section 5.
%-------------------------------------------------------------------------

\section{System Model and Problem Formulation}

\begin{figure}[!t]
\centering
\includegraphics[width=3in]{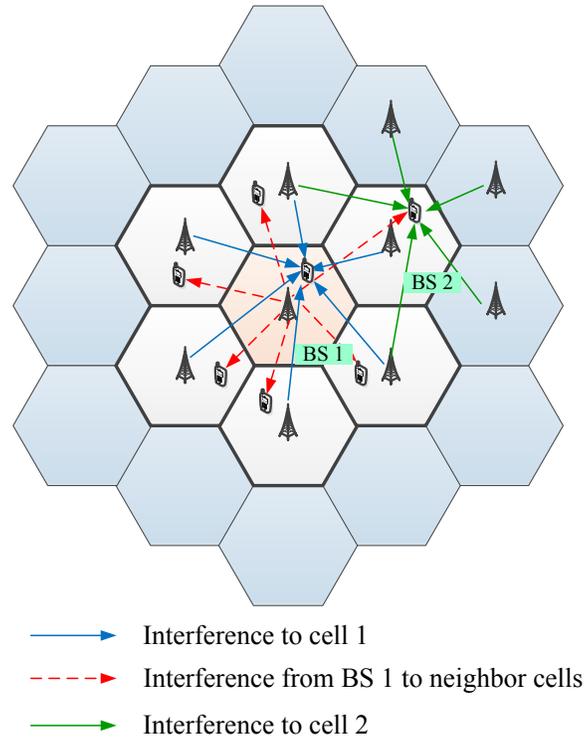}
\caption{System model} \label{system_model.eps}
\end{figure}

In this paper, we investigate a downlink OFDMA multi-cell network
with $M$ BSs and $N$ subcarriers. The BSs are connected by
high-speed fiber. Each user is served by only one BS. In addition,
each subcarrier is exclusively assigned to one user. As shown in
Fig.\ref{system_model.eps}, because of the openness of wireless
channels the users in one cell will inevitably receive the
interference signals from the neighboring BSs, especially the users
at the edge of a cell. This fact becomes the most distinguished
factor limiting wireless system performance. As a result,
interference coordination is of great importance and attracts a lot
of attention. Simply speaking, reducing transmit power will
definitely reduce mutual interference and save valuable resources to
realize green communications. However, with decreasing transmit
power meanwhile the throughput also decreases. It is not surprising
that in a complicated interference scenario there is hardly
closed-form optimal solutions. Due to the fact that the considered
optimization problem is in nature a multi-objective optimization
problem, it is impossible to define optimality. There definitely
exists some tradeoff between transmit powers at BSs and the whole
system performance within this awkward scenario and this is the
focus of our work. In view of this, relying on multiple-objective
optimization theory we want to formulate an optimization problem
with multiple objective functions and aim to derive a set of the
optimal solutions. The set will reflect different design preferences
and include several well-known special cases.

In our design for each BS $m$ the following two performance utility
functions are considered simultaneously when it optimizes its
resource allocation.

\noindent 1) Power consumption of the BS;

\noindent 2) Throughput contribution of the BS to the whole network.

\noindent It should be highlighted that the second performance
utility function consists of two parts. One is its own throughput
which is naturally a positive contribution to the whole network. The
other is the interference it caused, which is a negative
contribution as interference will decrease other terminal's
throughput. In the following, we will discuss the formulation of the
throughput contribution in detail.

\subsection*{2.1~~Throughput contribution of the BS to the network}

To measure the throughput contribution of the BS $m$ to the network,
pricing mechanism
\cite{MIMO,EW,beamforming,Beamformer,Non_Separable} is adopted in
our work. Specifically, the pricing-based multi-cell power
allocation game can be formulated as
\begin{equation}\label{game}
\mathcal{G}=\{\mathcal{M},\{\bm p_m\}_{m\in
\mathcal{M}},\{\tilde{U}_m\}_{m\in \mathcal{M}}\},
\end{equation}
where the involved elements are defined as follows \\
$\bullet$~\textbf{Player set}: The set of BSs is denoted by
$\mathcal{M}=\{1,2,\ldots,M\}$. \\
$\bullet$~\textbf{Strategy set}: The allocated power vector $\{\bm
p_1,\ldots,\bm p_M\}$ is defined as
\begin{equation}
\bm p_m=\{[p_m^{[1]},\ldots,p_m^{[N]}]^\textrm{T}:\sum_{n\in
\mathcal{N}}p_m^{[n]}\leq P_{\textrm{max}}\}.
\end{equation}\\
In the previous formulation, $p_m^{[n]}$ is the power allocated to
subcarrier $n$. $\mathcal{N}=\{1,2,\ldots,N\}$ is the set of
subcarriers with the
total number $N$, and $P_{\textrm{max}}$ is the maximum transmit power of each BS.\\
$\bullet$~\textbf{Payoff function set}: The payoff function set is
denoted by $\{\tilde{U}_1,\ldots,\tilde{U}_M\}$, where ${\tilde
{U}}_m$ is a function of the power allocation of all involved BSs.
To highlight this fact, in the following, we can write ${\tilde
{U}}_m=\tilde{U}_m(\bm p_m,\bm p_{-m})$ where ${\bm{p}}_m$ is the
power allocation of BS $m$ and ${\bm{p}}_{-m}$ is the power
allocations of the other BSs except BS $m$ with $\bm p_{-m}=\{\bm
p_{1},\ldots,\bm p_{m-1},\bm p_{m+1},\ldots,\bm p_{M}\}$.
Additionally all the power allocation vectors are assumed to be
known by all BSs.

In order to theoretically analyze the throughput contributions to
the whole network, we first investigate the throughput of BS $m$ on
subchannel $n$, which can be expressed as
\begin{equation}
\begin{split}
U_m^{[n]}
&=\log_2\left(1+\frac{{\left|h_{m,m}^{[n]}\right|}^2p_m^{[n]}}{\sigma^2+\underset{j\in
\mathcal{M}\setminus m}\sum {\left|h_{j,m}^{[n]}\right|}^2
p_j^{[n]}}\right)\\
& =\log_2\left(1+\frac{{\left|h_{m,m}^{[n]}\right|}
^2p_m^{[n]}}{\sigma^2+I_m^{[n]}}\right) \label{interference}
\end{split}
\end{equation}
where \DDcom{$h_{m,j}^{[n]},~\forall j,n$ denotes the complex
channel gain between BS $j$ and the user who is served by BS $m$ on
subchannel $n$}, $\sigma^2$ is the noise variance and $I_m^{[n]}$
denotes the interference term.

From (\ref{interference}) it can be seen that the relationship
between $U_m^{[n]}$ and $I_m^{[n]}$ is nonlinear. To overcome this
problem a linear model called pricing mechanism \cite{OFDMA} is
exploited to represent the total cost BS $m$ needs to pay to the
system when it applies $p_m^{[n]}$ on subchannel $n$.
\begin{equation}\label{total_cost}
\sum_{j\in \mathcal{M}\setminus
m}\pi_j^{[n]}{\left|h_{m,j}^{[n]}\right|}^2p_m^{[n]}
\end{equation}where $\pi_m^{[n]}$ is the interference pricing rate and is defined by \cite{OFDMA}
\begin{equation}\label{cost}
\pi_j^{[n]}=-\frac{\partial U_j^{[n]}}{\partial I_j^{[n]}}.
\end{equation}
Need to notice that the pricing rate requires to be updated after a
BS updates its power. Because only the updated pricing rate makes
sense for the neighboring BSs.

Summing up the costs in (\ref{total_cost}) across all the
subchannels $n\in \mathcal{N}$, the first utility function is
formulated as
\begin{equation}\label{utility_cost}
\tilde{U}_m(\bm p_m,\bm p_{-m})=\underset{n\in \mathcal{N}}\sum
            U_m^{[n]}-\underbrace{\underset{n\in
\mathcal{N}}\sum\underset{j\in \mathcal{M}\setminus
m}\sum\pi_j^{[n]}{\left|h_{m,j}^{[n]}\right|}^2p_m^{[n]}}_{\triangleq C(\bm p_m,\bm p_{-m})}\\
\end{equation}
where the term $C(\bm p_m,\bm p_{-m})$ can be interpreted to be the
cost of BS $m$ that should be paid to the system. As a result,
$\tilde{U}_m(\bm p_m,\bm p_{-m})$ is exploited to measure the
throughput contribution of the BS $m$ to the network.

\noindent\textbf{Remark}: The BSs receive channel state information
(CSI) from their serving users and exchange interference price as
well as a small portion of CSI information with neighboring BSs.
Different from the widely used non-linear Gauss-Seidel algorithm,
the BSs can make their decisions simultaneously with the pricing
mechanism.

\subsection*{2.2~~Problem Formulation}
According to the previous discussions, for the resource optimization
the following two objective functions should be minimized
simultaneously
\begin{align}
&f_1(\bm p_m)=-\underset{n\in \mathcal{N}}\sum
U_m^{[n]}+\underset{n\in \mathcal{N}}\sum\underset{j\in
\mathcal{M}\setminus m}\sum \pi_j^{[n]}{\left|h_{m,j}^{[n]}\right|}^2p_m^{[n]},\\
&f_2(\bm p_m)=\underset{n\in \mathcal{N}}\sum p_m^{[n]}
\end{align}based on which the multi-objective optimization problem (MOP) for each BS
$m$ can be formulated as
\begin{equation}\label{MOP}
\begin{split}
\textrm{MOP:}~~~~~&\underset{\bm
p_m}{\mathop{\textmd{min}}}~~~{\bm{f}}(\bm
p_m)=\left[\begin{array}{cccc}
             f_1(\bm p_m)\\
             f_2(\bm p_m)\\
  \end{array}\right]\\
& \textrm{s.t.}~~g_j(\bm p_m)\geq 0,~~j=1,\ldots,(N+1),\\
\end{split}
\end{equation}where the power constraints are defined as
\begin{equation}
\begin{aligned}
&g_j(\bm p_m)=p_m^{[j]},~~~~~~~~~~~~~~~~~j=1,\ldots,N,\\
&g_j(\bm p_m)=P_{\rm{max}}-\underset{n\in \mathcal{N}}\sum p_m^{[n]},~~j=N+1.\\
\end{aligned}
\end{equation}
As each BS solves MOP individually, in the following $\bm p_m$ is
replaced by ${\bm p}$ for notational simplicity.

Generally speaking, MOPs are usually much more difficult to solve
than its single objective counterpart. The difficulty in solving MOP
(\ref{MOP}) comes from its multi-objective functions and a common
logic to remove this difficulty relies on scalarization
\cite{Survey}. It should be pointed out that in the existing
wireless research, a common logic for scalarization is to replace
the original multiple objectives by their linear weighted sum.
Unfortunately, this kind of scalarization has the following
drawbacks. $\\$ \noindent (1) If the considered optimization problem
is nonconvex, the traditional scalarization will incur some losses.
In specifc, the set of final solutions may not be the Pareto optimal
set \cite{mopsp}. $\\$ \noindent (2) When searching the optimal
points, in the traditional scalarization method the adjustment of
weighting factors may make the objective value change nonuniformly.
As the optimal solutions are unknown, if the objective function
changes sharply some important points are not obtained
\cite{drawbacks}. $\\$ \noindent (3) The following operations after
the traditional scalarization are complicated for nonlinear MOP.
There are two reasons. First, linear weighted operation cannot
change the nonlinear nature of the objective function. Second, the
adaptive updating strategy for the weighting factors are largely
open. In most of the works, the weighting vectors are determined by
the importance among these objectives
\cite{weighted}\cite{weighted1}\cite{weighted2}. However, it is
difficult to find reasonable weighting vectors because there are
usually no direct relationship between weighting vectors and
objective functions, especially the objectives have different
physical meanings.$\\$ To overcome such weakness, from the
perspective of operational mathematics, another famous scalarization
technique named Pascoletti and Serafini Scalarization is much
preferred. In the following, a resource allocation algorithm is
proposed based on Pascoletti and Serafini Scalarization. To the best
of our knowledge, it is the first attempt to take advantage of
Pascoletti and Serafini Scalarization to design resource allocation
algorithm for multi-cell networks.

\section{The Proposed Algorithm Based on Multi-Objective Optimization Theory}

As there are two objective functions, the first question is how to
define the optimality or what kind of solutions we want to achieve.
Referring to the optimal solution, the following properties are
desired to be met. For a given constant power consumption, the
throughput contribution of the BS is the maximum one and on the
other hand for a fixed throughput contribution, the consumed power
is the minimum. From the viewpoint of multi-objective optimization,
this kind of solutions are named as K-minimal points \cite{mopsp}.

In Pascoletti and Serafini method, a new set of control parameters
equivalent to the weighting vectors are introduced. By varying the
parameters, the whole solution set of MOP (\ref{MOP}) can be found.
Thus in the following, we will first discuss the relationship
between MOP and the Pascoletti and Serafini scalarization problem
and then study how the new set of parameters affect minimal
solutions in subsection 3.1. And further in subsection 3.2, we will
study how to adaptively control the parameters such that the whole
solution set can be obtained.

\subsection*{3.1~~Pascoletti and Serafini Scalarization}

\begin{figure}[!t]
\centering
\includegraphics[width=4.4in]{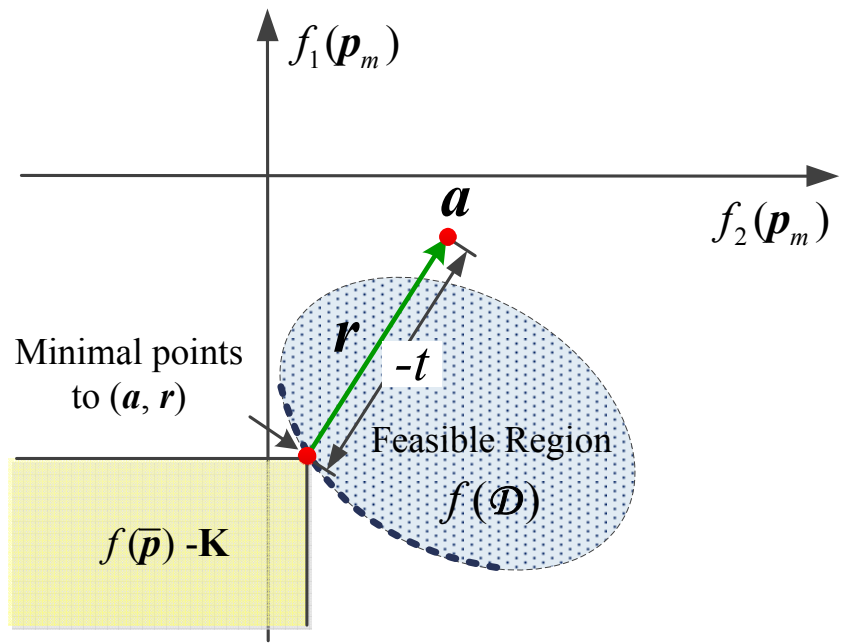}
\caption{Explanation of scalarization} \label{explain.eps}
\end{figure}

As discussed previously, we will choose a scalarization scheme named
Pascoletti and Serafini scalarization to transform the considered
multi-objective optimization problem into a single-objective
optimization problem (SP). Using the Pascoletti and Serafini method,
MOP (\ref{MOP}) is equivalent to the following SP,
\begin{equation}\label{(SP(a,r))}
\begin{split}
\textmd{SP:}~~~~~~\underset{t,\bm p\in
\mathcal{D}}{\mathop{\textmd{min}}}~~~&~~~~~~t\\
~~~\textmd{s.t.}~~~&\bm{a}+t\bm{r}-{\bm{f}}(\bm p)\in \textrm{\textbf{K}},\\
&g_j(\bm p)\geq 0,~~~j=1,\ldots,(N+1),\\
\end{split}
\end{equation}where the parameters are defined as $\bm a=[a_1 ~a_2]^\textrm{T} \in \mathbb{R}^2$ and
$\bm r=[r_1~ r_2]^\textrm{T} \in \mathbb{R}_+^2$. In addition,
$\textrm{\textbf{K}}=\mathbb{R}_+^2$ is the closed pointed convex
cone. The parameter $\bm a$ can be interpreted as a reference point
and the parameter $\bm r$ as a direction \cite{Helbig} as shown in
Fig. \ref{explain.eps}. For a reasonable $\bm r$ and $\bm a$ in the
coordinates, it is always possible to find a minimal $t$ ($<0$) and
then a K-minimal solution $\bar{\bm p}$ corresponding to this $(\bm
a,\bm r)$.

It has been proved in \cite{mopsp} that SP has all key properties a
scalarization method should have for determining minimal solutions
of MOP. If $(\bar{t},\bar{\bm p})$ is a minimal solution of SP, then
$\bar{\bm p}$ is a weakly K-minimal solution of MOP at least.
Besides, by varying $(\bm a,\bm r)$ all K-minimal points of MOP can
be found which are also the solutions of SP. It is obvious that
$(\bm{a},\bm{r})$ are the key parameters to SP and different
$(\bm{a},\bm{r})$ leads to different solutions to SP. As there is
more than one parameter in the transformation procedure, the
motivation is how to reduce the number of parameters and simplify
the computation for the optimal solutions.

\noindent \underline{\textbf{Conclusion 1:}} When the parameter $\bm
r$ is fixed, i.e., $\bm r=\bm r^0$, all K-minimal points can still
be found by varying the parameter $\bm a$.

\textsl{Proof:} The proof has been given in Section 6.1.

As a result, our attention is restricted  to the relationships
between parameter $\bm a$ and the minimal solution $\bm p$ as well
as ${\bm{f}}(\bm p)$ assuming $\bm r$ is constant. Regarding this,
we denote the SP w.r.t. the parameter $\bm{a}$ by
$\textrm{SP}(\bm{a})$ and regard $\bm p$ as a function of $\bm a$,
denoted by ${\bm p}(\bm a)$. Considering the perturbed parameters
$\bm a$ as
\begin{equation}\label{aa}
\bm a\approx \bm a^0 +s\bm v
\end{equation}
where $s \in \mathbb{R}_+$ is the distance between $\bm a$ and $\bm
a^0$, and $\bm v\in \mathbb{R}^{2}$ is the direction in which $\bm
a$ changes.

Substituting (\ref{aa}) into (\ref{(SP(a,r))}), the resulting
$\textrm{SP}({\bm a})$ becomes
\begin{equation}\label{(SP_modified)}
\begin{split}
\textrm{SP}({\bm a}):~~~~\underset{t,\bm p}{\mathop{\textmd{min}}}~~~&~~~~~~t\\
~~~\textmd{s.t.}~~~&(\bm a^0+s\bm v)+t\bm r^0-{\bm{f}}({\bm p})\geq \textbf{0}_2,\\
&{\bm{g}}(\bm p)\geq \textbf{0}_{(N+1)}.\\
\end{split}
\end{equation}

To find the whole solution set, we embark on  a reference problem
$\textrm{SP}(\bm{a}^0)$ and assume its minimal solution $(t^0(\bm
a^0),\bm p^0(\bm a^0))$ as well as the Lagrange multipliers
$(\bm{\mu}^0,\bm{\beta}^0)$ have already been obtained, where
$\bm{\mu}^0$ is the Lagrange multipliers to the first constraint of
(\ref{(SP_modified)}) and $\bm{\beta}^0$ to the second. Based on the
reference point, we can find the next solution $\bm p(\bm a)$ and
thus $\bm f(\bm p(\bm a))$ in the neighborhood of $\bm f(\bm p^0)$
with the help of directional derivatives. Then iteratively
implementing the process until the whole solution set is achieved.
To realize this, the relationship between $\bm a$ and $\bm p(\bm a)$
as well as $\bm f(\bm p(\bm a))$ should be investigated first.

Assume that a variation of the parameters $\bm a$ in one direction
only, i.e., $\bm v$ is fixed and the solutions of
(\ref{(SP_modified)}) only depend on the parameter $s$. Then, we can
regard all the terms $(t, \bm p, \bm{\mu}, \bm{\beta})$ as a
function of $s$, i.e., $\left(t(s), \bm p(s), \bm{\mu}(s),
\bm{\beta}(s)\right)$. For clarity, we use $(t, \bm p, \bm{\mu},
\bm{\beta})$ in the following. The Lagrange function of
(\ref{(SP_modified)}) is first given by:
\begin{equation}\label{Lagrange}
\mathcal{L}(t, \bm p, \bm{\mu},
\bm{\beta})=t-\sum_{i=1}^M{\mu}_i\left[(a_i^0+sv_i)+tr_i^0-f_i({\bm
p})\right]-\sum_{j=1}^{N+1}{\beta}_jg_j(\bm{p})
\end{equation}

For a solution in nonlinear programming \textrm{SP}($s$)
(\ref{(SP_modified)}) to be optimal, Karush-Kuhn-Tucker (KKT)
conditions are necessary:
\begin{equation}\label{KKT}
\begin{split}
\nabla_{t}\mathcal{L}(t, \bm p, \bm{\mu}, \bm{\beta})&=1-\sum_{i=1}^M{\mu}_ir_i^0=0\\
\nabla_{{\bm p}}\mathcal{L}(t, \bm p, \bm{\mu},
\bm{\beta})&=\sum_{i=1}^M{\mu}_i\nabla_{{\bm p}}f_i({\bm
p})-\sum_{j=1}^{N+1}{\beta}_j\nabla_{{\bm p}}g_j({\bm
p})=0\\
\mu_i\left((a_i^0+sv_i)+tr_i^0-f_i({\bm p})\right)&=0,~~\mu_i\geq0,~~~~~\forall i\\
\beta_jg_j({\bm p})&=0,~~\beta_j\geq0,~~~~~\forall j
\end{split}
\end{equation}

These nonlinear equations can be solved with the help of directional
derivatives, which is derived in Section 6.2.

With the solution $(\bar{t},\bar{\bm p}, \bar{\bm \mu},\bar{\bm
\beta})$ obtained by solving (\ref{system equations1}), the minimal
solution of \textrm{SP}($s$) can be attained by
\begin{equation}\label{s1}
\left(\begin{array}{cc}
             t \\
             {\bm p} \\
             \bm \mu \\
             \bm \beta \\
  \end{array}\right)\approx\left(\begin{array}{cc}
             t^0\\
             {\bm p}^0\\
             \bm \mu^0\\
             \bm \beta^0\\
  \end{array}\right)+s\cdot\left(\begin{array}{cc}
             \bar{t}\\
             \bar{{\bm p}}\\
             \bar{\bm \mu}\\
             \bar{\bm \beta}\\
  \end{array}\right)
\end{equation}
So far, the relationship between the minimal solution $({t},{\bm p},
{\bm \mu},{\bm \beta})$ and $s$ is obtained. And in Conclusion 2, we
will show how $s$ affect the minimal points $\bm f({\bm p})$.

\noindent \underline{\textbf{Conclusion 2:}} Since $s$ in (\ref{s1})
can be defined very small ($s\rightarrow 0^+$), the minimal points
$\bm f({\bm p})$ can be approximated using the first-order Taylor
approximation
\begin{equation}\label{approximation}
\bm f( {\bm p})\approx \bm f( {\bm p^0})+s\cdot \bm v+ s(\nabla_{\bm
a}\bar{\tau}^{\delta}(\bm a^0)^\textrm{T}\bm v)\bm r,
\end{equation}
where $\nabla_{\bm a}\bar{\tau}^{\delta}(\bm a^0)=-\bm \mu^0$ is the
derivative of the local minimal value function $\bar{\tau}^{\delta}$
in the point $\bm a^0$.

\textsl{Proof:} The proof can be found in Section 6.3.

Thus far, the function relationship among the minimal solution $\bm
p$, the minimal point $\bm f(\bm p)$ and $\bm a$ can be clearly
seen. Based on the results in (\ref{s1}) and (\ref{approximation}),
we then care how to adaptively control the parameter $\bm a$ such
that all the minimal solution $\bm p$ and the minimal point $\bm
f(\bm p)$ can be obtained instead of artificially modifying $\bm a$.

\subsection*{3.2~~Adaptive control of parameter $\bm a$}

In the following, a procedure is developed to achieve the whole
minimal solution set of the MOP by controlling the choice of the
parameter $\bm a$. First, it is necessary to give the set from which
the parameter $\bm a$ is chosen.

\noindent \textbf{\underline{Theorem 1:}} Define a hyperplane
\begin{equation}
\textrm{\textbf{H}}=\{\bm y\in \mathbb{R}^2|\bm b^\mathrm{T}\bm
y=\beta\},
\end{equation}
with $\bm b\in \mathbb{R}^2$, $\beta\in \{0,1\}$, and $\bm
b^\mathrm{T}\bm r\neq 0$. It is sufficient to get the efficient set
by varying the parameter $\bm a\in \textrm{\textbf{H}}$. Further, it
is shown that a subset $\bm a\in \textrm{\textbf{H}}^a\subset
\textrm{\textbf{H}}$ for
\begin{equation}\label{subset}
\textrm{\textbf{H}}^a=\{\bm y\in \mathbb{R}^2|\bm y=\lambda \bar{\bm
a}^1+(1-\lambda)\bar{\bm a}^2,~~\lambda\in[0,1]\},
\end{equation}
is also sufficient. Where $\bar{\bm a}^1\in \textrm{\textbf{H}}$ and
$\bar{\bm a}^2\in \textrm{\textbf{H}}$ are given by
\begin{equation}
\bar{\bm a}^i:=\bm f(\bar{\bm p}^i)-\bar{t}^i\bm r
~~~\textrm{with}~~\bar{t}^i:=\frac{\bm{b}^\textrm{T}\bm f(\bar{\bm
p}^i)-\beta}{\bm b^\mathrm{T} \bm r},~~~i=1,2.
\end{equation}
with $\bar{\bm p}^1$ the minimal solution of the scalar-valued
problem $\underset{\bm p\in \mathcal{D}}{\rm{min}}~~f_1(\bm p)$ and
$\bar{\bm p}^2$ the minimal solution of $\underset{\bm p\in
\mathcal{D}}{\rm{min}}~~f_2(\bm p)$. Without loss of generality, we
assume $\bar{\bm a}^1$ is smaller than $\bar{\bm a}^2$ on the first
dimension, i.e., $\bar{a}_1^1<\bar{a}_1^2$.

\textsl{Proof:} The proof can be found in our work \cite{ours}.

With the stricter set $\textrm{\textbf{H}}^a$ in (\ref{subset}) from
which $\bm a$ should be chosen, we now want to determine the
parameters $\bm a^0, \bm a^1, \bm a^2,\ldots$ adaptively (starting
with $\bm a^0=\bar{\bm a}^1$) such that the related minimal points
$\bm f(\bm p(\bm a^i)),~i=0,1,2,\ldots,$ gained by solving
$\textrm{SP}(\bm a^i)$ for $i=0,1,2,\ldots,$ have the equal distance
$\alpha>0$, i.e.,
\begin{equation}\label{aaaaa}
\|\bm f(\bm p(\bm a^0))-\bm f(\bm p(\bm a^1))\|=\alpha,
\end{equation}
should be satisfied for any neighboring $\bm a^i$ and $\bm a^{i+1}$.
This metric is applied to avoid the loss of some important points
caused by the sharply change of an objective function. In the
following, we replace $\bm f(\bm p(\bm a^0))$ with $\bm f(\bm p^0)$
for clarity. The advantage of choosing a predefined distance
$\alpha>0$ between two neighbor points can be seen from
Fig.\ref{explain.eps}. With the evenly distributed points, more
accurate information about the relationship between the objectives
would be known.

Then, we aim at finding a direction $\bm v$ and a scalar $s$ such
that $\bm a^1:=\bm a^0+s\bm v$ satisfies (\ref{aaaaa}). Based on
(\ref{approximation}), we have
\begin{equation}
\begin{split}
\alpha&=\|\bm f(\bm p^0)-\bm f(\bm p^1)\|\\
&\approx \|\bm f(\bm p^0)-(\bm f(\bm p^0)+s\bm v+s(-(\bm
\mu^0)^\textrm{T}\bm v)\bm
r)\| \\
&=|s|~\|\bm v+(-(\bm \mu^0)^\textrm{T}\bm v)\bm
r)\| .\\
\end{split}
\end{equation}
Then, the stepsize $s$ and direction $\bm v$ of parameter $\bm a$
can be chosen by
\begin{equation}\label{s_v}
\begin{split}
s^0&=\frac{\alpha}{\|\bm v+(-(\bm \mu^0)^\textrm{T}\bm v)\bm
r)\|},\\
\bm v&=\bm{a}^1-\bm{a}^0.\\
\end{split}
\end{equation}
which leads to the generalized formula of $\bm a$,
\begin{equation}\label{a_expression}
\bm a^{i+1}=\bm a^i + s^i\bm v=\bm a^i+\frac{\alpha}{\|\bm v+(-(\bm
\mu^i)^\textrm{T}\bm v)\bm r)\|}\bm v.
\end{equation}

Based on (\ref{a_expression}), an algorithm is developed to
adaptively solve $\textrm{SP}(\bm a^i)$ with $\bar{\bm a}^1\leq \bm
a^i\leq\bar{\bm a}^2$, whose solutions constitute the solution set.
This algorithm is referred to as adaptive parameter control (APC)
algorithm and is shown in \textbf{Algorithm 1}.

\begin{algorithm}
\caption{Adaptive Parameter Control Algorithm}

\noindent \hangafter=1 \setlength{\hangindent}{4em}
\textbf{Input:~~}Choose $\bm r=(r_1,r_2)^\textrm{T}\in
\mathbb{R}_+^2$ with $r_1>0$, and predefine $\alpha>0$ between the
neighboring two points. The hyperplane is chosen
as\\$~~~~~~~~~~~~\textrm{\textbf{H}}=\{\bm y\in \mathbb{R}^2|\bm
b^\textrm{T}\bm
y=\beta\}$, \\
where $\bm b\in \mathbb{R}^2$ and $\bm b^\textrm{T}\bm r\neq 0$,
$\beta \in \{0,1\}$. Given $M^1\in \mathbb{R}$
with\\
$~~~~~~~~M^1> \underset{\bm p\in \mathcal{D}}{\max}~f_2(\bm
p)-\underset{\bm p\in
\mathcal{D}}{\min}~f_1(\bm p)\frac{r_2}{r_1}$.\\

\noindent \hangafter=1 \setlength{\hangindent}{4em}
\textbf{Step~1:~~}Finding the minimal solution $(\tilde{t}^1,{\bm
p}^1)$ and Lagrange multiplier $\bm \mu^1\in \mathbb{R}_+^2$ of
$\textrm{SP}(\tilde{\bm a}^1)$ with $\tilde{\bm
a}^1=(0,M^1)^\textrm{T}$. Calculate\\
$~~~~~~t^1:=\frac{\bm b^\textrm{T}\bm f(\bm p^1)-\beta}{\bm
b^\textrm{T}\bm r}$ and $\bm
a^1:=\bm f(\bm p^1)-t^1\bm r$.\\
Set $l:=1$.\\

\noindent \hangafter=1 \setlength{\hangindent}{4em}
\textbf{Step~2:~~}Finding the minimal solution $\bm p^E$ of
$\underset{\bm p\in \mathcal{D}}{\min}~f_2(\bm
p)$ and set\\
$~~~~~t^E:=\frac{\bm b^\textrm{T}\bm f(\bm p^E)-\beta}{\bm
b^\textrm{T}\bm r}$ and $\bm
a^E:=\bm f(\bm p^E)-t^E\bm r$.\\
Let $\bm v:=\bm a^E-\bm a^1$.

\noindent \hangafter=1 \setlength{\hangindent}{4em}
\textbf{Step~3:~~}Update $\bm a^{l+1}$ by\\
$~~~\bm a^{l+1}:=\bm a^{l}+\frac{\alpha}{\|\bm v+(-(\bm
\mu^l)^\textrm{T}\bm v)\bm r)\|}\cdot \bm v$.

\noindent \hangafter=1 \setlength{\hangindent}{4em}
\textbf{Step~4:~~}Set $l:=l+1$.\\
$\bullet$ \textbf{If} $\bm a^l=\bm a^1+\rho \bm v$ for a $\rho \in
[0,1]$, find minimal solution $(t^l,\bm p^l)$ and Lagrange
multiplier $\bm \mu^l$ by solving $\textrm{SP}(\bm a^l)$, and go to step 3.\\
$\bullet$ \textbf{Else} stop.

\noindent \hangafter=1 \setlength{\hangindent}{4em}
\textbf{Output:~~}Set $\mathbf{\mathcal{A}}=\{\bm p^1,\ldots,\bm
p^{l-1},\bm p^E\}$ is the minimal solution set of MOP. Set
$\mathbf{\mathcal{B}}=\{\bm f(\bm p^1),\ldots,\bm f(\bm p^{l-1}),\bm
f(\bm p^E)\}$ is an approximation of the minimal points.

\end{algorithm}

In the \textbf{Input} Step, we arbitrarily choose a $\tilde{\bm
a}^1=(0,M^1)^\textrm{T}$ with $M^1> \underset{\bm p\in
\mathcal{D}}{\max}~f_2(\bm p)-\underset{\bm p\in
\mathcal{D}}{\min}~f_1(\bm p)\frac{r_2}{r_1}$ at the initial for the
purpose of engineering practice. Actually, the optimization problem
$\textrm{SP}(\tilde{\bm a}^1,\bm r)$ with such a $\tilde{\bm
a}^1=(0,M^1)^\textrm{T}$ in Step 1 is
\begin{equation}\label{(SP(a1,r))}
\begin{split}
&\underset{\bm p\in \mathcal{D}}{\min}~~~~~~~~t~\\
&~\textmd{s.t.}~~~~~~~~~tr_1-f_1(\bm
p)\geq 0,\\
&~~~~~~M^1+tr_2-f_2(\bm p)\geq 0,\\
&~~~~~~~~~~~~~~t\in \mathbb{R}.
\end{split}
\end{equation}
Therefore $t\geq \frac{f_1(\bm p)}{r_1}$ holds for any feasible
point $(t,\bm p)$, and $M^1+tr_2-f_2(\bm p)>(f_2(\bm p)-f_1(\bm
p)\frac{r_2}{r_1})+\frac{f_1(\bm p)}{r_1}r_2-f_2(\bm p)=0$ is also
satisfied. Therefore, (\ref{(SP(a1,r))}) can be replaced by
$\underset{\bm p\in \mathcal{D}}{\min}~~~\frac{f_1(\bm p)}{r_1}$
which equals to $\underset{\bm p\in
\mathcal{D}}{\textrm{min}}{f_1(\bm p)}$. However, there is no need
to find the exact solution of $\underset{\bm p\in
\mathcal{D}}{\textrm{min}}{f_1(\bm p)}$ in practice, because we can
always find the starting point $\bm a^1$ even if the initial
$\tilde{\bm a}^1$ is roughly given. In addition,
$\textrm{SP}(\tilde{\bm a}^1)$ and $\textrm{SP}({\bm a}^1)$ have the
same minimal solution ${\bm p}^1$ and the same Lagrange multiplier
$\bm \mu^1$, which will be demonstrated in \textbf{Theorem 2}.

\noindent \textbf{\underline{Theorem 2:}} Assume $(\bar{t},\bar{\bm
p})$ is a minimal solution of $\textrm{SP}(\bm a,\bm r)$ with
Lagrange multiplier $\bm \mu$ for arbitrary $\bm a\in \mathbb{R}^2$
and $\bm r\in \mathbb{R}^2$ with $\bm b^\mathrm{T}\bm r\neq 0$.
Surely there could be a $\bar{\bm k}\in \textrm{\textbf{K}}$ with
\begin{equation}
\bm a+\bar{t}\bm r-\bm f(\bar{\bm p})=\bar{\bm k}.
\end{equation}
Further there could be a $\bm a' \in \textrm{\textbf{H}}$ and $t'\in
\mathbb{R}$ such that $(t',\bar{\bm p})$ is a minimal solution of
$\textrm{SP}(\bm a',\bm r)$ with
\begin{equation}
\bm a'+t'\bm r-\bm f(\bar{\bm p})=\bm 0_2.
\end{equation}
In addition $\bm \mu$ is also Lagrange multiplier to the point
$(t',\bar{\bm p})$ for $\textrm{SP}({\bm a}',\bm r)$.

\textsl{Proof:} The proof can be found in our work \cite{ours}.

In Step 1 and Step 2, the subset $\textrm{\textbf{H}}^a$ of the
hyperplane and a direction $\bm v$ with $\bm a^1+s\bm v\in
\textrm{\textbf{H}}$ are determined. And then in Step 3 and Step 4,
the APC algorithm producing an approximation of the efficient set is
done.

\section{Simulation Results}

In this section, the performance the proposed APC algorithm is
assessed. A multi-carrier multi-user OFDM-based system is employed
with $M=19$ BSs as shown in Fig.\ref{system_model.eps}. There are 64
users in the simulation model and each user is served using a
randomly chosen subcarrier. The number of subcarriers is $N=64$ with
the system bandwidth 10 MHz. The distance between adjacent BSs is
1000 m. Reyleigh fast fading is considered, and large scale path
loss is modeled as $\textrm{PL} = 128.1 + 37.6\lg(d)$ \cite{3GPP},
where $d$ is the distance in kilometers.

As our work is studied based on the function in
(\ref{utility_cost}), we choose it as a comparison to the proposed
APC algorithm. This algorithm applies pricing mechanism, and aims at
only maximizing the transmission rate contribution of the BS to the
network, thus we denote it by ``pricing mechanism'' in the
simulations. The solutions of the equal power allocation and utility
maximization are also simulated as the reference points. Here,
utility maximization is that each BS tries to maximize his own
transmission rate ignoring the interference it may cause to other
cells and does not care how much power they will use to achieve the
maximum transmission rate. While the equal power allocation scheme
is that the total power needed in the APC method is uniformly
allocated to all subcarriers. For the initialization, we assume that
the total 30 watts energy is evenly distributed across all the
subcarriers.

\begin{figure}[!t]
\centering
\includegraphics[width=4.4in]{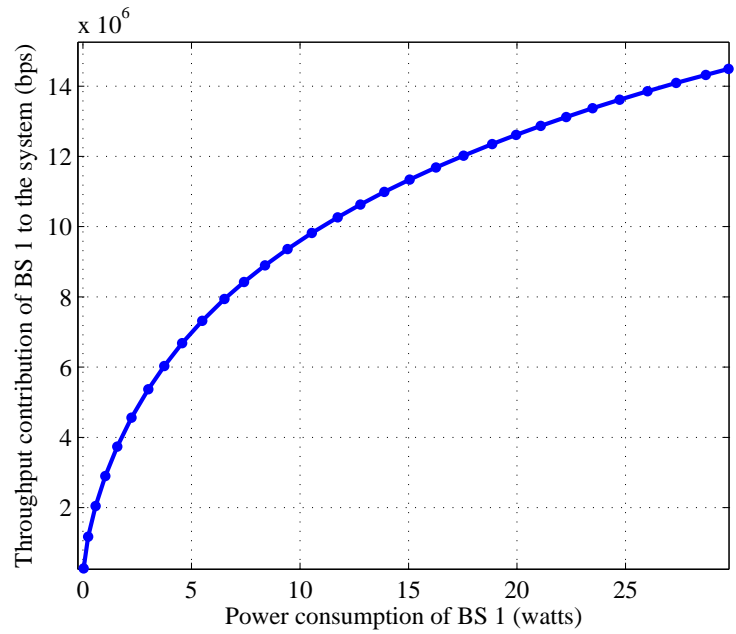}
\caption{Relationship between power consumption and throughput
contribution of the BS to the network} \label{power_throughput}
\end{figure}

Fig.\ref{power_throughput} shows the relationship between power
consumption and the throughput contribution of BS 1 to the network,
which is obtained by the APC method. This curve is the boundary of
the efficient set. For each given power consumption, the
corresponding throughput contribution of BS 1 to the network is the
maximum. And for the contribution of BS 1 to the network, the
corresponding power consumption is the minimum be required. When the
consumed power is greater than 20 watts, the increase of the network
throughput brought by the BS is small. Thus the extra 10 watts power
can be viewed as inefficient.

\begin{figure}[!t]
\centering
\includegraphics[width=4.4in]{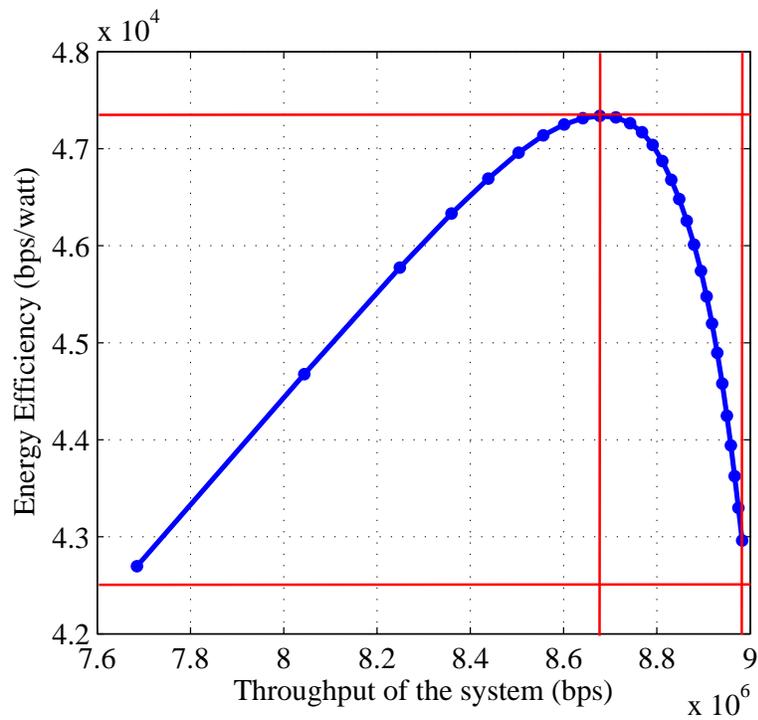}
\caption{Tradeoff between energy efficiency and throughput of the
system} \label{energy_efficiency}
\end{figure}

Fig.\ref{energy_efficiency} gives a parabola-like relationship
between energy efficiency and throughput of the system. Energy
efficiency is defined by the fraction where the numerator is the
throughput of the system and the denominator is the total power
consumption. As our goal is to find an energy-efficient power
allocation scheme which can also guarantee a relatively high system
throughput, we pay close attention to the interesting local area. If
the system agrees to reduce its transmission rate from 8.98 Mbps to
8.68 Mbps, energy efficiency will increase sharply to 47 kbps/watt,
which is much greater than the energy efficiency at 8.98 Mbps.

\begin{figure}[!t]
\centering
\includegraphics[width=5in]{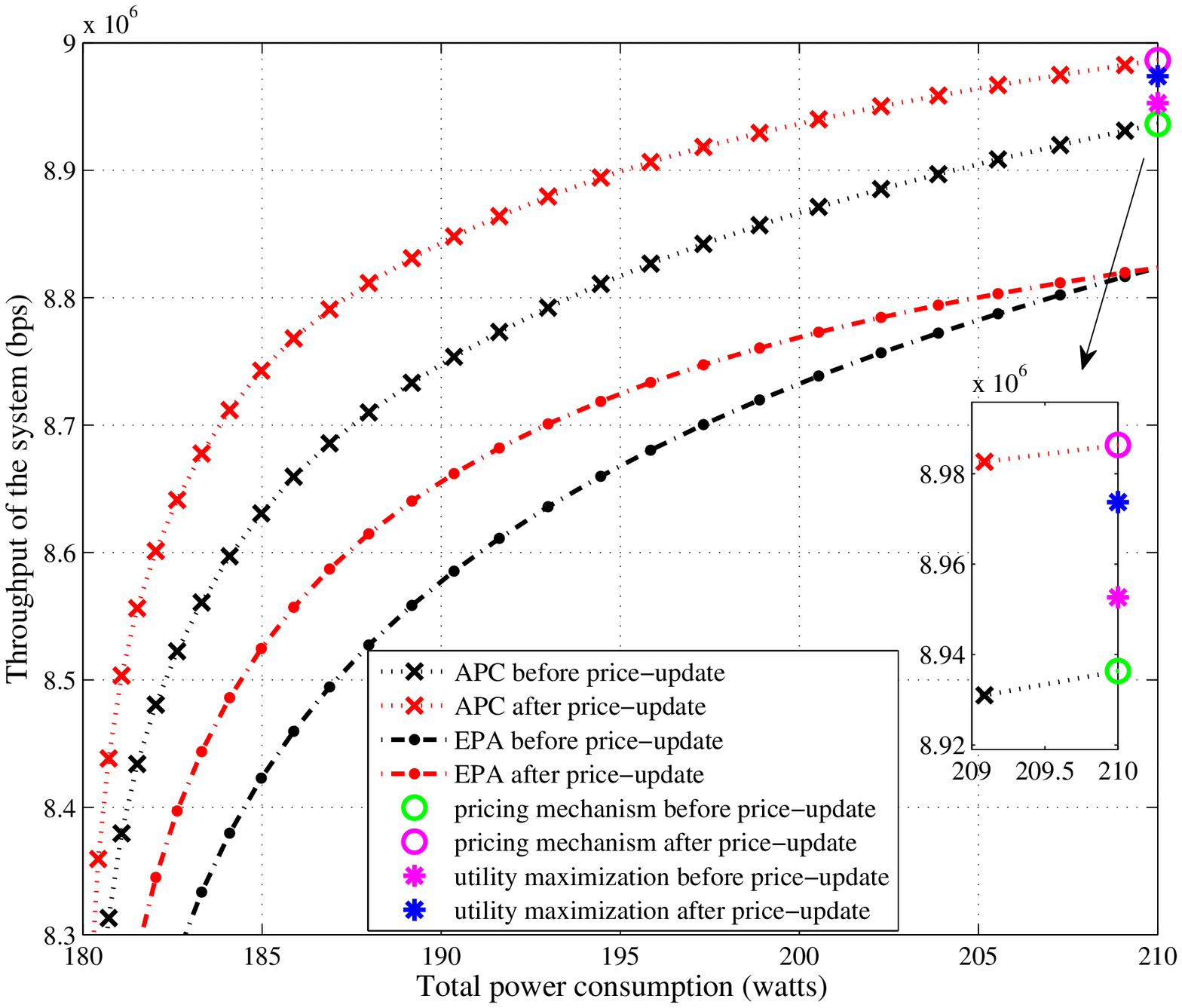}
\caption{Comparison between different schemes in view of throughput
of the system and total power consumption} \label{result}
\end{figure}

Fig.\ref{result} shows the comparison between different power
allocation schemes. All the curves are obtained by varying the power
consumption of BS 1 and fixing other BSs' power. It can be seen that
the solutions obtained by solving the pricing mechanism-based
problem are always on the APC curves, no matter whether the
interference pricing rate is updated or not. This result verifies
the comprehensiveness of the proposed APC method. Though it seems
that the solution to the utility maximization is better than the
solutions to pricing mechanism-based problem as well as APC problem
before update the pricing rate, its actual outcome is not so.
Because when we recalculate the system throughput after updating the
pricing rates, the solutions obtained by the pricing mechanism-based
scheme and APC method enjoy much more benefits than the greedy
utility maximization scheme. The reason is that the negative
contribution of BS 1 under utility maximization is great for it
introduces more interference to users in neighboring cells. A final
note about this figure is that all the curves are obtained by
varying the power consumption of BS 1 and fixing other BSs' power,
and the power allocation scheme adopted by the neighboring BSs is
EPA. As a result, there are no significant differences among these
considered schemes since only one player changed his strategy during
this process.

\section{Conclusions}

In this paper, from the perspective of multi-objective optimization
theory an energy-efficient power allocation scheme was developed for
interference-limited multi-cell network.  A bi-objective
optimization problem was first formulated based on the pricing
mechanism. Then using the method of Pascoletti and Serafini
scalarization, the relationship between the varying parameters and
minimal solutions has been discovered, and an adaptive algorithm was
developed to achieve tradeoff between the two objectives. As a
result, all the solutions on the boundary of the efficient solution
set can be computed which are best in terms of both energy
efficiency and inter-cell interference mitigation. Finally, the
performance of
the algorithm was demonstrated in the simulations.\\

\section{Acknowledgments}

This work was supported in part by the National Natural Science
Foundation of China under Grant 61101130, and the Excellent Young
Scholar Research Funding of Beijing Institute of Technology under
Grant 2013CX04038.

\section{Appendices}

\subsection*{6.1~~Proof of Conclusion 1}

Let $\bar{\bm p}$ be a K-minimal solution of the MOP, and set $\bm
a=\bm f(\bar{\bm p})$ and choose $\bm r\in
\textrm{\textbf{K}}\setminus \{\bm 0_2\}$ arbitrarily. Then the
point $(0,\bar{\bm p})$ is a feasible and also minimal solution for
$\textrm{SP}(\bm a,\bm r)$. Otherwise there will be a $t'<0$ and
$\bm p'\in \mathcal{D}$ with $(t',\bm p')$ feasible for
$\textrm{SP}(\bm a,\bm r)$, and a $\bm k'\in \textrm{\textbf{K}}$
with $\bm a+t'\bm r-\bm f(\bm p')=\bm k'$. With these observations,
we have
\[
\bm f(\bar{\bm p})=\bm f(\bm p')+\bm k'-t'\bm r\in \bm f(\bm
p')+\textrm{\textbf{K}}.\] Because $\bar{\bm p}$ is a K-minimal
solution, it can be concluded that $\bm f(\bar{\bm p})=\bm f(\bm
p')$ and thus $\bm k'=t'\bm r$. Since the cone $\textrm{\textbf{K}}$
is pointed, $\bm k'\in \textrm{\textbf{K}}$ and $t'\bm r\in
-\textrm{\textbf{K}}$, which leads to $\bm k'=t'\bm r=\bm 0_2$.
However, it is contradict to $t'<0$ and $\bm r\neq \bm 0_2$. Thus it
can be concluded that if $\bar{\bm p}$ is a minimal solution of the
MOP, then $(0,\bar{\bm p})$ is a minimal solution of
$\textrm{SP}(\bm a,\bm r)$ for the parameter $\bm a:=\bm f(\bar{\bm
p})$ and for arbitrarily $\bm r\in \textrm{\textbf{K}}\setminus
\{\bm 0_2\}$. Further, it can be inferred that all K-minimal points
can be found by varying the parameter $\bm a$ only.

\subsection*{6.2~~Derivation of the solution to (\ref{KKT})}

For the solution $(t, \bm p, \bm{\mu}, \bm{\beta})$ of
$\textrm{SP}(s)$ which is differentiable, the righthanded
derivatives $(\bar{t},\bar{{\bm p}},\bar{\bm \mu},\bar{{\bm
\beta}})$ can be written as
\begin{equation}\label{sys_equs}
\underset{\alpha\rightarrow 0^+}{\lim}\left(
\begin{array}{cccc}
             \frac{t-t^0}{\alpha}\\
              \frac{{\bm p}-{\bm p}^0}{\alpha}\\
              \frac{\bm \mu-\bm \mu^0}{\alpha}\\
              \frac{{\bm \beta}-{\bm \beta}^0}{\alpha}
\end{array}\right)=\left(
\begin{array}{cccc}
             \bar{t}\\
             \bar{{\bm p}}\\
             \bar{\bm \mu}\\
             \bar{{\bm \beta}}\\
\end{array}\right),
\end{equation}based on which it can also be inferred from Theorem 3 of \cite{131} that for
sufficiently small $\alpha$, there exists a unique continuous
function $(t, \bm p, \bm{\mu}, \bm{\beta})$ as the minimal solution
of $\textrm{SP}(s)$. Then we will find the solution $(t, \bm p,
\bm{\mu}, \bm{\beta})$ with the help of $(\bar{t},\bar{{\bm
p}},\bar{\bm \mu},\bar{{\bm \beta}})$ in (\ref{sys_equs}). Because
the KKT conditions (\ref{KKT}) is always satisfied, it follows that
the derivatives $(\bar{t},\bar{{\bm p}},\bar{\bm \mu},\bar{{\bm
\beta}})$ can be easily obtained by solving the following system of
inequalities and equations,
\begin{equation}\label{system equations1}
\begin{split}
-\sum_{i=1}^{M}\bar{\mu}_ir_i^0&=0,\\
\sum_{i=1}^{M}\mu_i^0\nabla_{{\bm p}}^2f_i({\bm p}^0)\bar{{\bm p}}-\sum_{i=1}^{N+1}\beta_i^0\nabla_{{\bm p}}^2&g_i({\bm p}^0)\bar{{\bm p}}\\
+\sum_{i=1}^{M}\bar{\mu}_i\nabla_{{\bm p}}f_i({\bm p}^0)-\sum_{i=1}^{N+1}\bar{\beta}_i^0\nabla_{{\bm p}}&g_i({\bm p}^0)=\bm 0_{N},\\
r_i^0\bar{t}-\nabla_{{\bm p}}f_i({\bm p}^0)^\mathrm{T}\bar{{\bm p}}&=-\bm v,~~~\forall i\in I^+,\\
r_i^0\bar{t}-\nabla_{{\bm p}}f_i({\bm p}^0)^\mathrm{T}\bar{{\bm p}}&\geq -\bm v,~~~\forall i\in I^0,\\
\bar{\mu}_i&\geq 0,~~~~~~\forall i\in I^0,\\
\bar{\mu}_i\big(r_i^0 \bar{t}-\nabla_{{\bm p}}f_i({\bm p}^0)^\mathrm{T}\bar{{\bm p}}+\bm v\big)&=0,~~~~~~\forall i\in I^0,\\
\bar{\mu}_i&=0,~~~~~~\forall i\in I^-.\\
\end{split}
\end{equation}
\begin{equation}\label{system equations2}
\begin{split}
\nabla_{{\bm p}}g_j({\bm p}^0)^\mathrm{T}\bar{{\bm p}}&=0,~~~~~~\forall j\in J^+,\\
\nabla_{{\bm p}}g_j({\bm p}^0)^\mathrm{T}\bar{{\bm p}}&\geq 0,~~~~~~\forall j\in J^0,\\
\bar{\beta}_j&\geq 0,~~~~~~\forall j\in J^0,\\
\bar{\beta}_j(\nabla_{{\bm p}}g_j({\bm p}^0)^\mathrm{T}\bar{{\bm p}})&= 0,~~~~~~\forall j\in J^0,\\
\bar{\beta}_j&= 0,~~~~~~\forall j\in J^-.\\
\end{split}
\end{equation}
where $I^+$, $I^0$, $I^-$ are the active, non-degenerate and
degenerate constraint sets, respectively
\begin{equation}\label{disjoint}
\begin{split}
&I^+:=\{i \in I\mid a_i^0+t^0r_i^0-f_i({\bm p}^0)=0,~~\mu_i^0>0\},\\
&I^0~:=\{i \in I\mid a_i^0+t^0r_i^0-f_i({\bm p}^0)=0,~~\mu_i^0=0\},\\
&I^-:=\{i \in I\mid a_i^0+t^0r_i^0-f_i({\bm p}^0)>0,~~\mu_i^0=0\},\\
\end{split}
\end{equation}
and $J^+$, $J^0$, $J^-$ are another disjoint sets:
\begin{equation}\label{disjoint}
\begin{split}
&J^+:=\{j \in J\mid g_j({\bm p}^0)=0,~~\beta_j^0>0\},\\
&J^0~:=\{j \in J\mid g_j({\bm p}^0)=0,~~\beta_j^0=0\},\\
&J^-:=\{j \in J\mid g_j({\bm p}^0)>0,~~\beta_j^0=0\}.\\
\end{split}
\end{equation}
With the derivatives $(\bar{t},\bar{{\bm p}},\bar{\bm \mu},\bar{{\bm
\beta}})$ obtained by solving (\ref{system equations1}) and
(\ref{system equations2}), the minimal solution of \textrm{SP}($s$)
can be attained from (\ref{sys_equs}), which equals
\begin{equation}
\left(\begin{array}{cc}
             t \\
             {\bm p} \\
             \bm \mu \\
             \bm \beta \\
  \end{array}\right)\approx\left(\begin{array}{cc}
             t^0\\
             {\bm p}^0\\
             \bm \mu^0\\
             \bm \beta^0\\
  \end{array}\right)+s\cdot\left(\begin{array}{cc}
             \bar{t}\\
             \bar{{\bm p}}\\
             \bar{\bm \mu}\\
             \bar{\bm \beta}\\
  \end{array}\right).
\end{equation}

\subsection*{6.3~~Proof of Conclusion 2}

In order to prove the following equation
\begin{equation}\label{f_prof}
\bm f( {\bm p})\approx \bm f( {\bm p^0})+s\cdot \bm v+ s(\nabla_{\bm
a}\bar{\tau}^{\delta}(\bm a^0)^\textrm{T}\bm v)\bm r,
\end{equation}
the definition of the local minimal value function
$\bar{\tau}^{\delta}$ should be given first, and then the derivative
of the local minimal value function in the point $\bm a^0$ should be
provided as well. The local minimal value function
$\tau^{\delta}:\mathbb{R}^2\rightarrow \mathbb{R}$ is defined by
\begin{equation}
{\tau}^{\delta}(\bm a):=\inf\{t\in \mathbb{R} | (t,\bm p)\in
\Sigma\bm a\cap B_{\delta}(t^0,\bm p^0)\},
\end{equation}
where $B_{\delta}(t^0,\bm p^0)$ is the closed ball with radius
$\delta>0$ around $\bm p^0$ and $\Sigma\bm a$ is the constraint set
of SP depending on $\bm a$:
\begin{equation}
\Sigma\bm a:=\{(t,\bm p)\in \mathbb{R}^{N+1} | \bm a+t\bm r-\bm f(
{\bm p})\geq \textbf{0}\}.
\end{equation}

Then the derivative can be achieved by
\begin{equation}\label{proof1}
\begin{split}
\nabla_{\bm a}&\tau^{\delta}(\bm a)=-\bm \mu(\bm a)\\
&-\sum_{i=1}^2\nabla_{\bm a}\bm \mu_i(\bm a)(a_i+t(\bm a)r_i-f_i(\bm p(\bm a))).\\
\end{split}
\end{equation}
Then replace $\bm \mu(\bm a^0)$ by $\bm \mu^0$, we obtain the
derivative in the point $\bm a^0$
\begin{equation}\label{proof2}
\begin{split}
\nabla_{\bm a}&\tau^{\delta}(\bm a^0)=-\bm \mu^0\\
&-\sum_{i\in I^+\bigcup I^0}\nabla_{\bm a}\bm \mu_i(\bm
a^0)\underbrace{(a_i^0+t^0r_i^0-f_i(\bm p^0))}_{=0} \\
&-\sum_{i\in I^-}\nabla_{\bm a}\bm \mu_i(\bm a^0)(a_i^0+t^0r_i^0-f_i(\bm p^0)). \\
\end{split}
\end{equation}
Due to the definition of $I^-$, $a_i^0+t^0r_i^0-f_i(\bm p^0)>0$ for
$i\in I^-$. Since $a_i+t(\bm a)r_i-f_i(\bm p(\bm a))$ is continuous
in $\bm a$, there exists a neighborhood $\mathcal{N}(\bm a^0)$ of
$\bm a^0$ so that for all $\bm a \in \mathcal{N}(\bm a^0)$ it
holds\\
\begin{equation}
a_i+t(\bm a)r_i-f_i(\bm p(\bm a))>0~~~~~\textrm{for}~~~i\in I^-.
\end{equation}
Then $\mu_i(\bm a)=0$ for all $\bm a\in \mathcal{N}(\bm a^0)$ and
$\nabla_{\bm a}\bm \mu_i(\bm a^0)=0$ for $i \in I^-$ can be
obtained. Thus, we get
\begin{equation}\label{proof3}
\nabla_{\bm a}\tau^{\delta}(\bm a^0)=-\bm \mu^0.
\end{equation}

Already obtained the derivative of the local minimal value function
$\nabla_{\bm a}\tau^{\delta}(\bm a^0)$, we then try to obtain the
result in (\ref{f_prof}) with the help of (\ref{proof3}).

Assume that we have already solved the problem $\textrm{SP}(\bm
a^0)$ for the parameters $\bm a^0$ with a minimal solution $(t^0,\bm
p^0)$ and Lagrange multiplier $\bm \mu^0$. Then by using
${\tau}^{\delta}(\bm a^0)={t}(\bm a^0)=t^0$, a first order Taylor
approximation of the local minimal value function to the
optimization problem $\textrm{SP}(\bm a)$ is derived,
\begin{equation}
{t}(\bm a)\approx t^0+\nabla_{\bm a} {\tau}^{\delta}(\bm
a^0)^\textrm{T}(\bm a-\bm a^0)
\end{equation}
Then the approximation for the K-minimal points of MOP dependent on
the parameter $\bm a$ is launched:
\begin{equation}
\begin{split}
\bm f( {\bm p}(\bm a))&=\bm a+t(\bm a)\bm r\\
&\approx \bm a^0+(\bm a-\bm a^0)+(t^0+\nabla_{\bm a}
{\tau}^{\delta}(\bm a^0)^\textrm{T}(\bm
a-\bm a^0))\bm r\\
&=\bm f(\bm p^0)+(\bm a-\bm a^0)+(\nabla_{\bm a} {\tau}^{\delta}(\bm
a^0)^\textrm{T}(\bm
a-\bm a^0))\bm r\\
&=\bm f(\bm p^0)+s\bm v+s(\nabla_{\bm a}\bar{\tau}^{\delta}(\bm
a^0)^\textrm{T}\bm v)\bm r.\\
\end{split}
\end{equation}

\subsection*{6.4~~Proof of Theorem 1}

First, we will prove that it is sufficient to vary the parameter
$\bm a$ on the hyperplane $\textrm{\textbf{H}}=\{\bm y\in
\mathbb{R}^2|\bm b^\textrm{T}\bm y=\beta\}$. Assume $\bar{\bm p}$ is
K-minimal for MOP. For the case that
\begin{equation}\label{aaa}
\bar{t}=\frac{\bm b^\textrm{T}\bm f(\bar{\bm p})-\beta}{\bm
b^\textrm{T} \bm r}~~~\textrm{and}~~~\bm a=\bm f(\bar{\bm
p})-\bar{t}\bm r
\end{equation}
with arbitrarily $\bm r\in \textrm{\textbf{K}}$ and $\bm
b^\textrm{T}\bm r\neq 0$, we have $\bm a\in \textrm{\textbf{H}}$ and
$(\bar{t},\bar{\bm p})$ is feasible for SP($\bm a$). If
$(\bar{t},\bar{\bm p})$ is not a minimal solution of SP($\bm a$),
then there could be another $t'<\bar{t}$, $\bm p'\in \mathcal{D}$
and $\bm k'\in \textrm{\textbf{K}}$ with
\begin{equation}
\bm a+t'\bm r-\bm f(\bm p')=\bm k'.
\end{equation}
Replace $\bm a$ with (\ref{aaa}),
\begin{equation}
\bm f(\bar{\bm p})=\bm f(\bm p')+\bm
k'+\underset{>0}{\underbrace{(\bar{t}-t')}}\underset{\in
\textrm{\textbf{K}} }{\underbrace{\bm r}},
\end{equation}
then it can be concluded that $\bm f(\bar{\bm p})\in \bm f(\bm
p')+\textrm{\textbf{K}}$ for $\bm p'\in \mathcal{D}$, which is
contradict to the definition of $\bar{\bm p}$ K-minimal. Thus
$(\bar{t},\bar{\bm p})$ is a minimal solution of SP($\bm a$). So
far, we have proved that it is sufficient to vary the parameter $\bm
a$ on the hyperplane $\textrm{\textbf{H}}$. Further, we will show
that a subset $\textrm{\textbf{H}}^a\subset \textrm{\textbf{H}}$ is
also sufficient to get the efficient set.

Assume $\bar{\bm p}^1$ is a minimal solution of (\ref{min_f1})
\begin{equation}\label{min_f1}
\underset{\bm p\in \mathcal{D}}{\rm{min}}~~{\bm l^1}^{\textrm{T}}\bm
f(\bm p)
\end{equation}
and $\bar{\bm p}^2$ is a minimal solution of (\ref{min_f2})
\begin{equation}\label{min_f2}
\underset{\bm p\in \mathcal{D}}{\rm{min}}~~{\bm l^2}^{\textrm{T}}\bm
f(\bm p)
\end{equation}
where $\bm l^1=(1,0)$ and $\bm l^2=(0,1)$. The parameters $\bar{\bm
a}^1\in \textrm{\textbf{H}}$ and $\bar{\bm a}^2\in
\textrm{\textbf{H}}$ are given by
\begin{equation}
\bar{\bm a}^i:=\bm f(\bar{\bm p}^i)-\bar{t}^i\bm r
~~~\textrm{with}~~\bar{t}^i:=\frac{\bm{b}^\textrm{T}\bm f(\bar{\bm
p}^i)-\beta}{\bm b^\mathrm{T} \bm r},~~~i=1,2.
\end{equation}

Then, we consider the parameters $\bm a\in \textrm{\textbf{H}}^a$
with the set $\textrm{\textbf{H}}^a$ given by
\begin{equation}
\textrm{\textbf{H}}^a=\{\bm y\in \textrm{\textbf{H}}|\bm y=\lambda
\bar{\bm a}^1+(1-\lambda)\bar{\bm a}^2,~~\lambda\in[0,1]\}.
\end{equation}
It can be inferred from the assumption that $\bar{\bm a}^1,\bar{\bm
a}^2\in \textrm{\textbf{H}}^a\subset \textrm{\textbf{H}}$. For
simplicity, we assume $\bar{\bm a}^1$ is smaller than $\bar{\bm
a}^2$ on the first dimension, i.e., $\bar{a}_1^1<\bar{a}_1^2$.

For any feasible $\bar{\bm p}$, there exists a parameter $\bm a\in
\textrm{\textbf{H}}$ and a $\bar{t}\in \mathbb{R}$ given by
\begin{equation}
\bar{t}=\frac{\bm b^\textrm{T}\bm f(\bar{\bm p})-\beta}{\bm
b^\textrm{T} \bm r}~~~\textrm{and}~~~\bm a=\bm f(\bar{\bm
p})-\bar{t}\bm r
\end{equation}
so that $(\bar{t},\bar{\bm p})$ is a minimal solution of
$\textrm{SP}(\bm a)$. As $\bar{\bm p}^1$ and $\bar{\bm p}^2$ are
minimal solutions of (\ref{min_f1}) and (\ref{min_f2}), we have for
any feasible $\bar{\bm p}$,
\begin{equation}\label{relation}
{\bm l^1}^\textrm{T}\bm f(\bar{\bm p})\geq {\bm l^1}^\textrm{T}\bm
f(\bar{\bm p}^1)~~\textrm{and}~~{\bm l^2}^\textrm{T}\bm f(\bar{\bm
p})\geq {\bm l^2}^\textrm{T}\bm f(\bar{\bm p}^2).
\end{equation}
Suppose that ${\bm l^1}^\textrm{T}\bm f(\bar{\bm p})\geq {\bm
l^1}^{\textrm{T}}\bm f(\bar{\bm p}^2)$, as (\ref{relation}) always
holds, it can be concluded that $\bm f(\bar{\bm p})-\bm f(\bar{\bm
p}^2)\in \textrm{\textbf{K}}$, which is contradict to $\bar{\bm p}$
K-minimal. Thus, it can be shown that
\begin{equation}\label{relation1}
{\bm l^1}^\textrm{T}\bm f(\bar{\bm p}^1)\leq {\bm l^1}^\textrm{T}\bm
f(\bar{\bm p})\leq {\bm l^1}^\textrm{T}\bm f(\bar{\bm p}^2).
\end{equation}
In the same way, the following relation can also be achieved
\begin{equation}\label{relation2}
{\bm l^2}^\textrm{T}\bm f(\bar{\bm p}^2)\leq {\bm l^2}^\textrm{T}\bm
f(\bar{\bm p})\leq {\bm l^2}^\textrm{T}\bm f(\bar{\bm p}^1).
\end{equation}

With the obtained relations (\ref{relation1}) and (\ref{relation2}),
we then demonstrate that the parameter $\bm a$ lies on the segment
between the point $\bar{\bm a}^1$ and $\bar{\bm a}^2$, i.e., $\bm
a=\lambda \bar{\bm a}^1+(1-\lambda)\bar{\bm a}^2$ for a $\lambda\in
[0,1]$. Using the definition of $\bm a$, $\bar{\bm a}^1$ and
$\bar{\bm a}^2$, the following can be obtained:
\begin{equation}\label{fff}
\bm a=\bm f(\bar{\bm p})-\bar{t}\bm r=\lambda (\bm f(\bar{\bm
p}^1)-\bar{t}^1\bm r)+(1-\lambda)(\bm f(\bar{\bm p}^2)-\bar{t}^2\bm
r).
\end{equation}
reformulate (\ref{fff}) as
\begin{equation}\label{reformf}
\bm f(\bar{\bm p})=\lambda \bm f(\bar{\bm p}^1)+(1-\lambda)\bm
f(\bar{\bm p}^2)+(\bar{t}-\lambda \bar{t}^1-(1-\lambda)\bar{t}^2)\bm
r.
\end{equation}
Then we do a case differentiation for $\bar{t}-\lambda
\bar{t}^1-(1-\lambda)\bar{t}^2\geq 0$ and $\bar{t}-\lambda
\bar{t}^1-(1-\lambda)\bar{t}^2<0$ respectively.

For $\bar{t}-\lambda \bar{t}^1-(1-\lambda)\bar{t}^2\geq 0$, we first
divide the set of $\lambda$ into three parts, i.e., $\lambda<0$,
$0<\lambda<1$ and $\lambda>1$, then start by considering the case
that $\lambda<0$.
\begin{equation}
\begin{split}
&~~~~{\bm l^1}^{\textrm{T}}\bm f(\bar{\bm p})\\
&=\lambda {\bm l^1}^\textrm{T}\bm f(\bar{\bm p}^1)+(1-\lambda){\bm
l^1}^\textrm{T}\bm f(\bar{\bm p}^2)+\underbrace{(\bar{t}-\lambda
\bar{t}^1-(1-\lambda)\bar{t}^2)}_{\geq 0}\underbrace{{\bm
l^1}^\textrm{T}\bm
r}_{\geq 0} \\
&\geq \underbrace{\lambda}_{<0} \underbrace{{\bm l^1}^\textrm{T}\bm
f(\bar{\bm p}^1)}_{<{\bm l^1}^\textrm{T}\bm f(\bar{\bm
p}^2)}+(1-\lambda){\bm
l^1}^\textrm{T}\bm f(\bar{\bm p}^2) \\
&>\lambda {\bm l^1}^\textrm{T}\bm f(\bar{\bm p}^2)+(1-\lambda){\bm
l^1}^\textrm{T}\bm f(\bar{\bm p}^2)\\
&={\bm l^1}^\textrm{T}\bm f(\bar{\bm p}^2).
\end{split}
\end{equation}
If ${\bm l^1}^\textrm{T}\bm f(\bar{\bm p})>{\bm l^1}^\textrm{T}\bm
f(\bar{\bm p}^2)$ is satisfied together with ${\bm
l^2}^\textrm{T}\bm f(\bar{\bm p})>{\bm l^2}^\textrm{T}\bm f(\bar{\bm
p}^2)$, it can be concluded that $\bm f(\bar{\bm p})-\bm f(\bar{\bm
p}^2)\in \mathbf{K}$, which is contradicted to $\bar{\bm p}$
K-minimal. Therefore, (\ref{reformf}) is not satisfied for $\lambda
<0$. Then we consider the case that $\lambda>1$.
\begin{equation}
\begin{split}
&~~~~{\bm l^2}^\textrm{T}\bm f(\bar{\bm p})\\
&=\lambda {\bm l^2}^\textrm{T}\bm f(\bar{\bm p}^1)+(1-\lambda){\bm
l^2}^\textrm{T}\bm f(\bar{\bm p}^2)+\underbrace{(\bar{t}-\lambda
\bar{t}^1-(1-\lambda)\bar{t}^2)}_{\geq 0}\underbrace{{\bm
l^2}^\textrm{T}\bm
r}_{\geq 0} \\
&\geq \lambda{\bm l^2}^\textrm{T}\bm f(\bar{\bm
p}^1)+\underbrace{(1-\lambda)}_{<0} \underbrace{{\bm
l^2}^\textrm{T}f(\bar{\bm
p}^2)}_{<{\bm l^2}^\textrm{T}\bm f(\bar{\bm p}^1)}\\
&>{\bm l^2}^\textrm{T}\bm f(\bar{\bm p}^1).
\end{split}
\end{equation}
If ${\bm l^2}^\textrm{T}\bm f(\bar{\bm p})>{\bm l^2}^\textrm{T}\bm
f(\bar{\bm p}^1)$ is satisfied together with ${\bm
l^1}^\textrm{T}\bm f(\bar{\bm p})>{\bm l^1}^\textrm{T}\bm f(\bar{\bm
p}^1)$, it can be concluded that $\bm f(\bar{\bm p})-\bm f(\bar{\bm
p}^1)\in \mathbf{K}$, which is contradicted to $\bar{\bm p}$
K-minimal. Thus, (\ref{reformf}) is not satisfied for $\lambda
>1$. Therefore, it can be concluded that (\ref{reformf}) for the case $\bar{t}-\lambda \bar{t}^1-(1-\lambda)\bar{t}^2\geq 0$ can only be
satisfied for $\lambda \in[0,1]$.

Then we consider the case $\bar{t}-\lambda
\bar{t}^1-(1-\lambda)\bar{t}^2<0$. We first consider the case
$\lambda>1$,
\begin{equation}\label{lamnda1}
\begin{split}
&~~~~{\bm l^1}^\textrm{T}\bm f(\bar{\bm p})\\
&=\lambda {\bm l^1}^\textrm{T}\bm f(\bar{\bm p}^1)+(1-\lambda){\bm
l^1}^T\bm f(\bar{\bm p}^2)+\underbrace{(\bar{t}-\lambda
\bar{t}^1-(1-\lambda)\bar{t}^2)}_{<0}\underbrace{{\bm
l^1}^\textrm{T}\bm
r}_{\geq 0} \\
&\leq \lambda {\bm l^1}^\textrm{T}\bm f(\bar{\bm
p}^1)+(1-\lambda){\bm
l^1}^\textrm{T}\bm f(\bar{\bm p}^2).\\
\end{split}
\end{equation}
As ${\bm l^1}^\textrm{T}\bm f(\bar{\bm p})>{\bm l^1}^\textrm{T}\bm
f(\bar{\bm p}^1)$, (\ref{lamnda1}) can be reformulated as
\begin{equation}\label{lamnda1re}
\underbrace{(\lambda-1)}_{> 0}\big({\bm l^1}^\textrm{T}f(\bar{\bm
p}^1)-{\bm l^1}^\textrm{T}f(\bar{\bm p}^2)\big)>0,
\end{equation}
which is contradict to (\ref{relation}).

For the case $\lambda<0$, it can be obtained in the same way that
\begin{equation}\label{lamnda2}
\lambda {\bm l^2}^\textrm{T}\bm f(\bar{\bm p}^1)+(1-\lambda){\bm
l^2}^\textrm{T}\bm f(\bar{\bm p}^2)\geq {\bm l^2}^\textrm{T}\bm
f(\bar{\bm p})>{\bm l^2}^\textrm{T}\bm f(\bar{\bm p}^2)
\end{equation}
and further we have
\begin{equation}\label{lamnda2re}
\underbrace{(-\lambda)}_{> 0}\big({\bm l^2}^\textrm{T}\bm f(\bar{\bm
p}^2)-{\bm l^2}^\textrm{T}\bm f(\bar{\bm p}^1)\big)>0,
\end{equation}
which is also contradict to (\ref{relation}). Therefore, it can be
concluded that (\ref{reformf}) for the case $\bar{t}-\lambda
\bar{t}^1-(1-\lambda)\bar{t}^2<0$ can only be satisfied for $\lambda
\in[0,1]$.

Based on the previous results, the following conclusion can be drawn: For any
K-minimal solution $\bm p$ of MOP, there exists a parameter $\bm
a\in \textrm{\textbf{H}}^a$ and some $\bar{t}\in \mathbb{R}$ so that
$(\bar{t},\bar{\bm p})$ is
a minimal solution of $\textrm{SP}(\bm a)$.\\

\subsection*{6.5~~Proof of Theorem 2}

Defining the following auxiliary variables
\begin{align}
& t':=\frac{\bm b^\textrm{T}\bm f(\bar{\bm p})-\beta}{\bm
b^\textrm{T} \bm
r} \\
& \bm a':=\bm a+(\bar{t}-t')\bm r-\bar{\bm k}=\bm f(\bar{\bm
p})-t'\bm r,
\end{align} it is straightforward that $\bm a'\in \textrm{\textbf{H}}$ and $\bm a'+t'\bm r-\bm f(\bar{\bm
p})=\bm 0_2$. The point $(t',\bar{\bm p})$ is feasible for
$\textrm{SP}(\bm a')$ and it is also a minimal solution, because
otherwise there exists a feasible point $(\hat{t},\hat{\bm p})$ of
$\textrm{SP}(\bm a')$ with $\hat{t}<t'$ and some $\hat{k}\in
\mathbf{K}$ with
\begin{equation}
\bm a'+\hat{t}\bm r-\bm f(\hat{\bm p})=\hat{\bm k}
\end{equation}
together with the definition of $\bm a'$, it can be concluded that
\begin{equation}
\bm a+(\bar{t}-t'+\hat{t})\bm r-\bm f(\hat{\bm p})=\hat{\bm
k}+\bar{\bm k}\in \mathbf{K}.
\end{equation}
Hence, $(\bar{t}-t'+\hat{t},\hat{\bm p})$ is feasible for
$\textrm{SP}(\bm a)$ with $\bar{t}-t'+\hat{t}<\bar{t}$, which is in
contradiction to the minimality of $(\bar{t},\bar{\bm p})$ for
$\textrm{SP}(\bm a)$. Thus, $(t',\bar{\bm p})$ is also a minimal
solution of $\textrm{SP}(\bm a')$.

Then we demonstrate that the two scalar problems have the same
Lagrange multiplier $\bm \mu$. The Lagrange function $\mathcal{L}$
to the scalar optimization problem $\textrm{SP}(\bm a)$ related to
the MOP is given by
\begin{equation}
\mathcal{L}(t,\bm p,\bm \mu, \bm \beta, \bm a)=t-\bm
\mu^\textrm{T}(\bm a+t\bm r-\bm f(\bm p))- \bm \beta^\textrm{T}\bm
g(\bm p).
\end{equation}
If $\bm \mu$ is Lagrange multiplier to the point $(\bar{t},\bar{\bm
p})$, then it follows
\begin{equation}
\begin{split}
&\nabla_{(t,\bm p)}\mathcal{L}(\bar{t},\bar{\bm p},\bm \mu, \bm
\beta, \bm a)^\textrm{T}\left(\begin{array}{cc}
             t-\bar{t}\\
             \bm p-\bar{\bm p}\\
  \end{array}\right)\\
&=\left[\left(\begin{array}{cc}
             1\\
             0\\
  \end{array}\right)-\sum_{i=1}^{2}\mu_i\left(\begin{array}{cc}
             r_i\\
             -\nabla_{\bm p}f_i(\bar{\bm p})\\
  \end{array}\right)-\sum_{i=1}^{N+1}\beta_i\left(\begin{array}{cc}
             0\\
             \nabla_{\bm p}g_i(\bar{\bm p})\\
  \end{array}\right)\right]^\textrm{T}\left(\begin{array}{cc}
             t-\bar{t}\\
             \bm p-\bar{\bm p}\\
  \end{array}\right).
\end{split}
\end{equation}
Hence $1-\bm \mu^\textrm{T}\bm r=0$ and $(\bm
\mu^\textrm{T}\nabla_{\bm p}{\bm f(\bar{\bm p})})(\bm p-\bar{\bm
p})\geq 0$. Further we have $\bm \mu^\textrm{T}(\bm a+t\bm r-\bm
f(\bm p))=0$. For the minimal solution $(t',\bar{\bm p})$ of the
problem $\textrm{SP}(\bm a')$, it is
\begin{equation}
\bm a'+t'\bm r-\bm f(\bar{\bm p})=\bm 0_2,
\end{equation}
and thus $\bm \mu^\textrm{T}(\bm a'+t'\bm r-\bm f(\bar{\bm p}))=0$.
Together with the following equality
\begin{equation}
\nabla_{(t,\bm p)}\mathcal{L}(t',\bar{\bm p},\bm \mu, \bm \beta, \bm
a')=\nabla_{(t,\bm p)}\mathcal{L}(\bar{t},\bar{\bm p},\bm \mu, \bm
\beta, \bm a),
\end{equation}
we also have
\begin{equation}
\nabla_{(t,\bm p)}\mathcal{L}(t',\bar{\bm p},\bm \mu, \bm \beta, \bm
a')^\textrm{T}\left(\begin{array}{cc}
             t-t'\\
             \bm p-\bar{\bm p}\\
  \end{array}\right)\geq 0.
\end{equation}
Therefore, $\bm \mu$ is also Lagrange multiplier to the point
$(t',\bar{\bm p})$ for the problem $\textrm{SP}(\bm a')$.

\newpage

\begin{thebibliography}{18}

\bibitem{Boudreau}
Boudreau, G., Panicker, J., Ning Guo, Rui Chang, Neng Wang and
Vrzic, S.,
\newblock ``Interference coordination and cancellation for 4G networks,''
\newblock {\em IEEE Commun. Mag.}, 2009, \textbf{47}, (4), pp. 74--81

\bibitem{Fodor}
Fodor, G., Koutsimanis, C., R\'{a}cz, A., Reider, N., Simonsson, A.
and M\"{u}ller, W.,
\newblock ``Intercell interference coordination in OFDMA networks and in the 3GPP long
term evolution system,''
\newblock {\em J. Commun.}, 2009, \textbf{4}, pp. 445--453

\bibitem{FeiGao}
Gao, F., Zhang, R., Liang, Y.-C. and Wang, X.,
\newblock ``Design of Learning Based MIMO Cognitive Radio Systems,''
\newblock {\em IEEE Trans. Veh. Technol.}, 2010, \textbf{59}, (4), pp.
1707--1720

\bibitem{Lasaulce}
Lasaulce, S., Debbah M., Altman, E.,
\newblock ``Methodologies for analyzing equilibria in wireless games,''
\newblock {\em IEEE Signal Process. Mag.}, 2009, \textbf{26}, (5), pp. 41--52

\bibitem{Sung}
Sung, C.W., Wong, W.S.,
\newblock ``A Noncooperative Power Control Game for Multirate CDMA Data Networks,''
\newblock {\em IEEE Trans. on Wirel. Commun.}, 2003, \textbf{2}, (1), pp. 186--194

\bibitem{Menon}
Menon, R., MacKenzie, A.B., Hicks, J.E., Buehrer, R.M., Reed, J.H.,
\newblock `` Game-Theoretic Framework for Interference Avoidance,''
\newblock {\em IEEE Trans. Commun.}, 2009, \textbf{57}, (4), pp. 1087--1098

\bibitem{Liang}
Liang, L., Gang, F.,
\newblock `` A Game-Theoretic Framework for Interference Coordination in OFDMA
Relay Networks,''
\newblock {\em IEEE Trans. Veh. Technol.}, 2012, \textbf{61}, (1), pp. 321--332

\bibitem{Huang} L.~Huang, Y.~Zhou, X.~Han, Y.~Wang, M.~Qian
and J.~Shi,
\newblock ``Distributed Coverage Optimization for Small Cell Clusters using Game
Theory,''
\newblock {\em IEEE Wirel. Commun. and Network Conf.}, 2013, pp. 2289--2293

\bibitem{Schmidt}
Schmidt, D.A., Shi, C., Berry, R.A., Honig, M.L., Utschick, W.,
\newblock ``Distributed resource allocation schemes,''
\newblock {\em IEEE Signal Process. Mag.}, 2009, \textbf{26}, (5), pp. 53--63

\bibitem{Non_Separable}
Shi, C., Berry, R.A., and Honig, M.L.,
\newblock ``Distributed interference pricing for ofdm wireless networks with
  non-separable utilities,''
\newblock {\em 42nd Annual Conf. Information Sciences and Systems}, 2008, pp. 755--760

\bibitem{OFDMA}
Xu, W., Wang, X.,
\newblock ``Pricing-based distributed downlink beamforming in multi-cell ofdma networks,''
\newblock {\em IEEE J. Sel. Areas Commun.}, 2012, \textbf{30}, (9), pp. 1605--1613

\bibitem{MIMO}
Changxin S., Schmidt, D.A., Berry, R.A., Honig, M.L., Utschick, W.,
\newblock ``Distributed interference pricing for the mimo interference
  channel,''
\newblock {\em IEEE Int. Conf. Commun.}, 2009, pp. 1--5

\bibitem{EW}
Ho, Z.K.M., Kaynia, M., Gesbert, D.,
\newblock ``Distributed power control and beamforming on mimo interference
  channels,''
\newblock {\em European Wirel. Conf.}, 2010, pp. 654--660

\bibitem{beamforming}
Shi, C., Berry, R.A., Honig, M.L.,
\newblock ``Local interference pricing for distributed beamforming in mimo
  networks,''
\newblock {\em IEEE Military Commun. Conf.}, 2009, pp. 1--6

\bibitem{Beamformer}
Schmidt, D.A., Shi, C., Berry, R.A., Honig, M.L., Utschick, W.,
\newblock ``Distributed resource allocation schemes: Pricing algorithms for
  power control and beamformer design in interference networks,''
\newblock {\em IEEE Signal Process. Mag.}, 2009, \textbf{26}, (5), pp. 53--63

\bibitem{unified_approach}
Scutari, G., Palomar, D.P., Facchinei, F., Pang, J.S.,
\newblock ``Distributed dynamic pricing for mimo interfering multiuser systems:
  A unified approach,''
\newblock {\em Int. Conf. Network Games, Control and
  Optimization}, 2011, pp. 1--5

\bibitem{Fapojuwo} J.~Rao and A.~Fapojuwo,
\newblock ``A Survey of Energy Efficient Resource Management Techniques for Multicell Cellular Networks,''
\newblock {\em IEEE Commun. Surveys $\&$ Tutorials}, \textbf{PP}, (99), May 2013, pp
1--27

\bibitem{Choi} H.~Ji, Y.~Kim, S.~Choi, J.~Cho and J.~Lee,
\newblock ``Dynamic resource adaptation in beyond LTE-A TDD heterogeneous networks,''
\newblock {\em IEEE Int. Conf. Commun. Workshops (ICC)}, 2013, pp
133--137

\bibitem{mopsp}
Eichfelder, G.,
\newblock {\em Adaptive Scalarization Methods in Multiobjective Optimization},
\newblock {(Springer-Verlag Berlin Heidelberg, 2008)}

\bibitem{Survey}
Marler, R. T., Arora, J.S.,
\newblock ``Survey of multi-objective optimization methods for engineering,''
\newblock {\em Structural and Multidisciplinary Optimization}, 2004, \textbf{26}, pp. 369--395


\bibitem{drawbacks}
Das, I., Dennis, J. E.,
\newblock ``A closer look at drawbacks of minimizing weighted sums of objectives for pareto set generation in multicriteria optimization
problems,''
\newblock {\em Structural Optimization}, 1997, \textbf{14}, pp.
63--69

\bibitem{weighted}
Devarajan, R., Jha, S.C., Phuyal, U., Bhargava, V.K.,
\newblock ``Energy-Aware Resource Allocation for Cooperative Cellular Network
Using Multi-Objective Optimization Approach,''
\newblock {\em IEEE Trans. on Wirel. Commun.}, 2012, \textbf{5}, (11), pp. 1797--1807

\bibitem{weighted1}
He, C., Sheng, B., Zhu, P., You X., Li, G.Y.,
\newblock ``Energy- and Spectral-Efficiency Tradeoff for Distributed Antenna Systems with Proportional Fairness,''
\newblock {\em IEEE J. Sel. Areas Commun.}, 2013, \textbf{31}, (5), pp
894--902

\bibitem{weighted2}
Iosifidis, G., Koutsopoulos, I.,
\newblock ``Double Auction Mechanisms for Resource Allocation in Autonomous Networks,''
\newblock {\em IEEE J. Sel. Areas Commun.}, 2010, \textbf{28}, (1), pp
95--102

\bibitem{Helbig}
Helbig, S.,
\newblock ``An interactive algorithm for nonlinear vector optimization,''
\newblock {\em Applied Mathematics and Optimizaiton}, 1990, \textbf{22}, pp. 147--151

\bibitem{131}
Jittorntrum, K.,
\newblock ``Solution point differentiability without strict complementarity in
  nonlinear programming,''
\newblock {\em Math Program Study}, 1984, \textbf{21}, pp. 127--138

\bibitem{ours} Li, N., Fei, Z., Xing, C., Kuang, J.,
\newblock ``Adaptive Multi-objective Optimization for Energy Efficient Interference Coordination in Multi-Cell Networks,''
\newblock {\em ArXiv identifier:1308.4777}, 2013

\bibitem{3GPP}
3GPP TS 36.814,
\newblock ``Evolved universal terrestrial radio access (e-utra),''
\newblock {2010}

\end{thebibliography}
\end{document}